\documentclass{article}
\usepackage{amsfonts}
\usepackage{amsmath}

\newtheorem{theorem}{Theorem}

\newtheorem{definition}[theorem]{Definition}

\input{tcilatex}
\begin{document}

\title{Canonical variational completion of differential equations}
\author{Nicoleta Voicu \\
%EndAName
"Transilvania"\ University, Brasov, Romania \and Demeter Krupka \\
%EndAName
Lepage Research Institute, Czech Republic}
\date{}
\maketitle

\begin{abstract}
Given a non-variational system of differential equations, the simplest way
of turning it into a variational one is by adding a correction term. In the
paper, we propose a way of obtaining such a correction term, based on the
so-called Vainberg-Tonti Lagrangian, and present several applications in
general relativity and classical mechanics.
\end{abstract}



\bigskip

\section{Introduction}

For a given non--variational system of differential equations, there are
multiple ways of transforming it into a variational one - among these,
variational multipliers (or variational integrating factors), \cite{Anderson}%
, are maybe the most well known. Another possibility is to simply add a
correction term.

In the paper, we consider systems of ordinary or partial differential
equations - represented by \textit{source forms, }or\textit{\ source
tensors, }similar to Euler-Lagrange systems for extremals of integral
variational functionals in the calculus of variations. We propose a way of
obtaining such a correction term - which we call a variational completion,
as follows. Any ordinary or partial differential system can be expressed as
the vanishing of some source form $\varepsilon $ on sections of an
appropriate jet bundle. Further, to this source form, one can naturally
attach a Lagrangian $\lambda _{\varepsilon }$, called the Vainberg-Tonti
Lagrangian of $\varepsilon $, \cite{Krupka-VT-Lagrangian}; this Lagrangian
has the property that the difference%
\begin{equation}
\tau :=E(\lambda _{\varepsilon })-\varepsilon  \label{canonical_completion_1}
\end{equation}%
between its Euler-Lagrange form $E(\lambda _{\varepsilon })$ and $%
\varepsilon $ offers a measure of the non-variationality of $\varepsilon $.
Using $\tau $ in (\ref{canonical_completion_1}) as the correction term, the
system $\varepsilon +\tau =0$ becomes variational.

The method appears to have several interesting applications. We present here
three of them.

1) \textit{Einstein tensor obtained from variational completion of the Ricci
tensor. }Historically, the first variant of gravitational field equations
proposed by Einstein was:%
\begin{equation}
R_{ij}=\dfrac{8\pi \kappa }{c^{4}}T_{ij},  \label{incomplete_efe}
\end{equation}%
where: $R_{ij}$ is the Ricci tensor of a 4-dimensional Lorentzian manifold $%
(X,g),$ $T_{ij}$ is the energy-momentum tensor, while $\kappa $ and $c$ are
constants, \cite{Landau}. This variant correctly predicted some physical
facts, but failed to fulfil another request: local energy-momentum
conservation. This led Einstein to adding in the left hand side the
"correction term" $-\dfrac{1}{2}Rg_{ij}$ (by a reasoning based on Bianchi
identities), thus leading to the nowadays famous:%
\begin{equation}
R_{ij}-\dfrac{1}{2}Rg_{ij}=\dfrac{8\pi \kappa }{c^{4}}T_{ij}.  \label{efe}
\end{equation}%
\ The variational deduction of (\ref{efe}), due to Hilbert, relies on a
heuristic argument - simplicity. Hilbert chose to construct the action for
the left hand side using the "simplest scalar" (i.e., simplest differential
invariant) which can be constructed from the metric tensor and its
derivatives alone. Happily, the Euler-Lagrange expression ensuing from this
simplest scalar - which is the scalar curvature $R$ - coincides with the
left hand side of (\ref{efe}).

There is, still, another way of finding this correction term. Equation (\ref%
{incomplete_efe}) is not variational. Actually, the term which fails to be
variational is $R_{ij};$ in the paper, we prove that the Hilbert Lagrangian
is (up to multiplication by a non-essential constant), nothing else than the
Vainberg-Tonti Lagrangian corresponding to $R_{ij}.$ Accordingly, the
correction term $-\dfrac{1}{2}Rg_{ij}$ can be obtained from $R_{ij}$ as a
canonical variational completion.

2)\ \textit{Energy-momentum tensors}. In special relativity, energy-momentum
tensors are obtained by adding to the Noether current corresponding to the
invariance of the matter Lagrangian to space-time translations a
symmetrization term. The way of obtaining the symmetrization term is subject
to an old debate, \cite{Forger}, \cite{Gotay}. The canonical variational
completion method offers the possibility of recovering the expression of a
full, symmetric energy-momentum tensor from just one of its terms - e.g.,
from a non-symmetrized Noether current. In particular, the energy-momentum
tensor of the electromagnetic field can be obtained this way.

3)\ In \textit{classical mechanics}, equations of \textit{damped small
oscillations} are known to be non-variational. Without aiming to give a
general physical interpretation of the obtained correction term, we
determine the canonical variational completion of these equations.

In Sections 2 and 3, we briefly present some known notions and results to be
used in the following.

\section{Differential forms on jet bundles}

The mathematical background for a modern formulation of both field theory
and mechanics are fibered manifolds and their jet bundles.

Consider a fibered manifold $Y$ of dimension $m+n,$ with $n$-dimensional
base $X$ and projection $\pi :Y\rightarrow X.$ Fibered charts $(V,\psi )$, $%
\psi =(x^{i},y^{\sigma })$ on $Y$ induce the fibered charts $(V^{r},\psi
^{r}),$ $\psi ^{r}=(x^{i},y^{\sigma },y_{~j_{1}}^{\sigma
},...,y_{~j_{1}j_{2}...j_{r}}^{\sigma })$ on the $r$-jet prolongation $%
J^{r}Y $ of $Y$ and $(U,\phi ),$ $\phi =(x^{i})$ on $X.$ The manifold $%
J^{r}Y $ can be regarded as a fibered manifold in multiple ways, by means of
the projections: 
\begin{equation*}
\pi ^{r,s}:J^{r}Y\rightarrow J^{s}Y,~\ \ (x^{i},y^{\sigma
},y_{~j_{1}}^{\sigma },...,y_{~j_{1}j_{2}...j_{r}}^{\sigma })\mapsto
(x^{i},y^{\sigma },y_{~j_{1}}^{\sigma },...,y_{~j_{1}j_{2}...j_{s}}^{\sigma
}),
\end{equation*}%
where $r>s,$ $J^{0}Y:=Y$ and:%
\begin{equation*}
\pi ^{r}:J^{r}Y\rightarrow X.
\end{equation*}

The set of $\mathcal{C}^{\infty }$-smooth sections $\gamma :X\rightarrow Y$,
locally expressed by some functions $(x^{i})\mapsto \gamma
(x^{i})=(x^{i},y^{\sigma }(x^{i}))$ is denoted by $\Gamma (Y).$ Given a
section $\gamma \in \Gamma (Y),$ its prolongation to $J^{r}Y$ is: $%
J^{r}\gamma :(x^{i})\mapsto J^{r}\gamma (x^{i})=(x^{i},y^{\sigma
}(x),y_{~,j}^{\sigma }(x),...,y_{~,j_{1}j_{2}...j_{r}}^{\sigma }(x))$, where
the symbol $_{,j}$ denotes partial differentiation with respect to $x^{j}.$

In field theoretical applications, the coordinates $x^{i}$ play the role of
space-time coordinates, while $y^{\sigma }$ are "field" coordinates (to be
accurate, real fields are encoded in \textit{sections} $y^{\sigma
}=y^{\sigma }(x^{i})$). The case of mechanics is characterized by $\dim X=1;$
in this case, the coordinates on $J^{r}Y$ are usually denoted by $%
(t,q^{\sigma },\dot{q}^{\sigma },\ddot{q}^{\sigma },...,q^{(r)})$ and are
interpreted as:\ time, generalized coordinates, generalized velocity etc.

By $\Omega _{k}^{r}W,$ we denote the set of $k$-forms of order $r$ over an
open set $W\subset Y,$ i.e., the set of $k$-forms over the $r$-th
prolongation $J^{r}W\subset J^{r}Y$. In particular, $\mathcal{F}(W):=\Omega
_{0}^{r}W$ is the set of real-valued smooth functions over $J^{r}W.$

The subset of $\Omega _{k}^{r}W$ consisting of $k$-forms:%
\begin{equation}
\rho =\dfrac{1}{k!}a_{i_{1}i_{2}...i_{k}}dx^{i_{1}}\wedge dx^{i_{2}}\wedge
...\wedge dx^{i_{k}},  \label{horizontal_form}
\end{equation}%
(where $a_{i_{1}i_{2}...i_{k}},$ $k\leq n,$ are smooth functions of the
coordinates $x^{i},y^{\sigma },y_{~j_{1}}^{\sigma
},...,y_{~j_{1}j_{2}...j_{r}}^{\sigma }$) is called the set of ($\pi ^{r}$-)%
\textit{horizontal }$k$-\textit{forms }of order $r;$ similarly, one can
speak about $\pi ^{r,s}$-horizontal forms of order $r$ as forms generated by
exterior products of the differentials $dx^{i},dy^{\sigma
},...,dy_{~j_{1}...j_{s}}^{\sigma }.$

Examples of $\pi ^{r}$-horizontal forms are\ volume forms and Lagrangians.

For $X=\mathbb{R}^{n}$, the \textit{Euclidean volume form} is:%
\begin{equation}
\omega _{0}=dx^{1}\wedge dx^{2}\wedge ...\wedge dx^{n}.
\label{volume_noninvariant}
\end{equation}%
\bigskip On pseudo-Riemannian manifolds $(X,g_{ij})$, a coordinate-invariant
volume form is locally given by: $dV=\sqrt{\left\vert g\right\vert }\omega
_{0},$ where $g:=\det (g_{ij}).$

A \textit{Lagrangian} of order $r$ is defined as a $\pi ^{r}$-horizontal $n$%
-form\textit{\ }of order $r:$%
\begin{equation}
\lambda =\mathcal{L}\omega _{0},~\ \ \ \ \ \ \mathcal{L=L}(x^{i},y^{\sigma
},...,y_{i_{1}...i_{r}}^{\sigma }).  \label{general_Lagrangian}
\end{equation}

\bigskip

A form $\theta \in \Omega _{k}^{r}Y$ is a \textit{contact form} if it is
annihilated by all jets $J^{r}\gamma $ of sections $\gamma \in \Gamma (Y).$
Important examples are the \textit{basic contact 1-forms} on $J^{r}Y$
defined on a coordinate neighborhood by:%
\begin{eqnarray}
&&\omega ^{\sigma }=dy^{\sigma }-y_{~j}^{\sigma }dx^{j},~\ \ \omega
_{~i_{1}}^{\sigma }=dy_{~i_{1}}^{\sigma }-y_{~i_{1}j}^{\sigma }dx^{j},...
\label{contact_1-forms} \\
&&~\ \ \ \omega _{~i_{1}i_{2}...i_{r-1}}^{\sigma
}=dy_{~i_{1}i_{2}...i_{r-1}}^{\sigma }-y_{~i_{1}i_{2}...i_{r-1}j}^{\sigma
}dx^{j}.  \notag
\end{eqnarray}%
A differential form is called $p$-contact if it is generated by $p$-th
exterior powers of contact forms.

\section{Source forms and variationality conditions}

A \textit{source form\ }of order $r$ on a fibered manifold $Y,$ \cite%
{Krupka-natural-L}, is a $\pi ^{r,0}$-horizontal, 1-contact $(n+1)$-form on $%
J^{r}Y$. In local coordinates, any source form is expressed as:%
\begin{equation}
\varepsilon =\varepsilon _{\sigma }\omega ^{\sigma }\wedge \omega _{0},~\ \
\ \varepsilon _{\sigma }=\varepsilon _{\sigma }(x^{i},y^{\sigma
},y_{~i}^{\sigma },...,y_{~j_{1}...j_{r}}^{\sigma }).  \label{source_form}
\end{equation}%
The set of source forms of order at most $r$ over $Y$ is closed under
addition and under multiplication with functions $f\in \mathcal{F}(J^{r}Y).$

The most notable example of a source form is the \textit{Euler Lagrange form}
$E(\lambda )$ of a Lagrangian $\lambda =\mathcal{L}(x^{i},y^{\sigma
},...,y_{i_{1}...i_{r}}^{\sigma })\omega _{0}\in \Omega _{n}^{r}(Y):$%
\begin{eqnarray*}
&&E(\lambda ):=E_{\sigma }\omega ^{\sigma }\wedge \omega _{0},~~\ \ \  \\
E_{\sigma } &=&\dfrac{\partial \mathcal{L}}{\partial y^{\sigma }}-d_{k_{1}}%
\dfrac{\partial \mathcal{L}}{\partial y_{~k_{1}}^{\sigma }}%
+...+(-1)^{r}d_{k_{1}}...d_{k_{r}}\dfrac{\partial \mathcal{L}}{\partial
y_{~k_{1}...k_{r}}^{\sigma }}.
\end{eqnarray*}

A section $\gamma :X\rightarrow Y$ is critical for the Lagrangian $\lambda $
if and only if the $E(\lambda )$ is annihilated by the $r$-jet of $\gamma ,$
i.e., $E_{\sigma }(\lambda )\circ J^{r}\gamma =0,$ $\sigma =1,...,m.$

\bigskip

A source form $\varepsilon $ is called:

a)\ \textit{locally variational} if around any point of the fibered manifold 
$Y,$ there exists a local fibered chart $(V,\psi )$ and a Lagrangian $%
\lambda $ on some jet prolongation $V^{r}$ ($r\in \mathbb{N}$) of $V,$ such
that, on $V^{r},$ $\varepsilon =E(\lambda );$

b)\ \textit{globally variational} if there exists a Lagrangian $\lambda $ on
the whole manifold $Y$ such that $\varepsilon =E(\lambda ).$

Local variationality of a source form $\varepsilon =\varepsilon _{\sigma
}\omega ^{\sigma }\wedge \omega _{0}$ of order $r$ is equivalent to a
generalization of classical Helmholtz conditions, \cite%
{Krupka-local-structure-L}:%
\begin{equation}
H_{\sigma \nu }^{~~~j_{1}...j_{k}}(\varepsilon )=0,~\ \ k=0,...,r,
\label{helmholtz_conds}
\end{equation}%
where:

\begin{eqnarray}
&&H_{\sigma \nu }^{~~~j_{1}...j_{k}}(\varepsilon )=\dfrac{\partial
\varepsilon _{\sigma }}{\partial y_{~j_{1}...j_{k}}^{\nu }}-(-1)^{k}\dfrac{%
\partial \varepsilon _{\nu }}{\partial y_{~j_{1}...j_{k}}^{\sigma }}-
\label{helmholtz_expr} \\
&&-\underset{l=k+1}{\overset{r}{\sum }}%
(-1)^{l}(_{k}^{l})d_{i_{k+1}}d_{i_{k+2}}...d_{i_{l}}\dfrac{\partial
\varepsilon _{\nu }}{\partial y_{~j_{1}...j_{k}i_{k+1}...i_{l}}^{\sigma }} 
\notag
\end{eqnarray}%
locally describe the \textit{Helmholtz form }$H_{\varepsilon }=\dfrac{1}{2}%
\overset{r}{\underset{k=0}{\sum }}H_{\sigma \nu
}^{~~~j_{1}...j_{k}}(\varepsilon )\omega _{~j_{1}...j_{k}}^{\nu }\wedge
\omega ^{\sigma }\wedge \omega _{0}.$

\section{Canonical variational completion}

By \textit{variational completion} of a given source form $\varepsilon $ on $%
Y,$ we will mean any source form $\tau $ on $Y$ with the property that $%
\varepsilon +\tau $ is variational. Of course, one can speak about local and
about global variational completions.

In the following, we will only study local variational completions.

Clearly, every source form has infinitely many variational completions:
indeed, any Lagrangian $\lambda $ induces the completion $\tau :=E(\lambda
)-\varepsilon $. Thus, the question is how to choose the Lagrangian $\lambda 
$ in a meaningful way. In the following, we will try to give an answer to
this question.

\bigskip

Given an arbitrary source form $\varepsilon =\varepsilon _{\sigma }\omega
^{\sigma }\wedge \omega _{0}\in \Omega _{n+1}^{r}Y$ of order $r$, a local
Lagrangian attached to $\varepsilon $ is the \textit{Vainberg-Tonti
Lagrangian }$\lambda _{\varepsilon }=\mathcal{L}_{\varepsilon }\omega _{0},$ 
\cite{Krupka-VT-Lagrangian}, \cite{Grigore}, defined by:%
\begin{equation}
\mathcal{L}_{\varepsilon }(x^{i},y^{\sigma },...,y_{~j_{1}...j_{s}}^{\sigma
})=y^{\sigma }\overset{1}{\underset{0}{\int }}\varepsilon _{\sigma
}(x^{i},uy^{\sigma },...,uy_{~j_{1}...j_{s}}^{\sigma })du.
\label{Vainberg-Tonti_L}
\end{equation}

The Euler-Lagrange form $E(\lambda _{\varepsilon })=E_{\nu }\omega ^{\nu
}\wedge \omega _{0}$ of the Vainberg-Tonti Lagrangian $\lambda _{\varepsilon
}$ is given, \cite{Krupka-VT-Lagrangian}, by:%
\begin{equation*}
E_{\nu }=\varepsilon _{\nu }-\underset{0}{\overset{1}{\int }}u\{y^{\sigma
}(H_{\nu \sigma }\circ \chi _{u})+y_{~j}^{\sigma }(H_{\nu \sigma }^{\
~j}\circ \chi _{u})+...+y_{~j_{1}...j_{r}}^{\sigma }(H_{\nu \sigma
}^{~~j_{1}...j_{r}}\circ \chi _{u})\}du,
\end{equation*}%
where $\chi _{u}:J^{2r}Y\rightarrow J^{2r}Y$ denotes the homothety $%
(x^{i},y^{\sigma },y_{~j}^{\sigma },...,y_{~j_{1}...j_{2r}}^{\sigma
})\mapsto (x^{i},uy^{\sigma },uy_{~j}^{\sigma
},...,uy_{~j_{1}...j_{2r}}^{\sigma })$ and the coefficients $H_{\sigma \nu
}^{~\ ~j_{1}...j_{k}}$ are as in (\ref{helmholtz_conds}).

From (\ref{helmholtz_conds}), it follows that the coefficients $H_{\sigma
\nu }^{~~~j_{1}...j_{k}}$ above have the meaning of "obstructions from
variationality" of the source form $\varepsilon .$ In particular, if the
source form $\varepsilon $ is variational, then $E(\lambda _{\varepsilon
})=\varepsilon .$

\bigskip

It thus appears as natural

\begin{definition}
The canonical variational completion of a source form $\varepsilon \in
\Omega _{n+1}^{r}(Y),$ is the source form $\tau (\varepsilon )$ given by the
difference between the Euler-Lagrange form of the Vainberg--Tonti Lagrangian
of $\varepsilon $ and $\varepsilon $ itself:%
\begin{equation}
\tau (\varepsilon )=E(\lambda _{\varepsilon })-\varepsilon .
\label{canonical_completion}
\end{equation}
\end{definition}

The local coefficients $\tau _{\nu }$ of the canonical variational
completion $\tau (\varepsilon )=\tau _{\nu }\omega ^{\nu }\wedge \omega _{0}$
can be directly expressed in terms of the coefficients $H_{\nu \sigma
}^{~~~j_{1}...j_{k}}$: 
\begin{equation*}
\tau _{\nu }=-\underset{0}{\overset{1}{\int }}u\{y^{\sigma }(H_{\nu \sigma
}\circ \chi _{u})+y_{~j}^{\sigma }(H_{\nu \sigma }^{~~j}\circ \chi
_{u})+...+y_{~j_{1}...j_{r}}^{\sigma }(H_{\nu \sigma
}^{~~j_{1}...j_{r}}\circ \chi _{u})\}du.
\end{equation*}

\bigskip

\textbf{Remark. }Generally speaking, the Vainberg-Tonti Lagrangian and,
accordingly, the canonical variational completion of a source form of order $%
r,$ are of order $2r.$ Still, under certain conditions, \cite{Grigore}
(which are fulfilled by a large number of equations in physics), the
Vainberg-Tonti Lagrangian is actually equivalent to a Lagrangian of order $%
r. $

\section{Source forms in general relativity}

Consider a Lorentzian manifold $(X,g_{ij})$ of dimension 4, with local
charts $(U,\phi )$, $\phi =(x^{i})_{i=\overline{0,3}}$ and Levi-Civita
connection $\nabla .$ We denote by\ $R_{ij}$ the Ricci tensor of $\nabla $
and by $R=g^{ij}R_{ij},$ the scalar curvature.\ We assume in the following
that measurement units are chosen in such a way that $c=1.$ Indices of
tensors will be lowered or raised by means of the metric $g_{ij}$ and its
inverse $g^{ij}$.

Einstein field equations (\ref{efe})\ arise by varying with respect to the
metric tensor the Lagrangian $\lambda =\lambda _{g}+\lambda _{m},$ where:

\textit{i)} $\lambda _{g}=-\dfrac{1}{16\pi \kappa }R\sqrt{\left\vert
g\right\vert }\omega _{0}$ (with $\omega _{0}=dx^{0}\wedge dx^{1}\wedge
dx^{2}\wedge dx^{3}$) is the Hilbert Lagrangian;

\textit{ii)} the \textit{matter Lagrangian} $\lambda _{m}=L_{m}\sqrt{%
\left\vert g\right\vert }\omega _{0},~$is given by a differential invariant $%
L_{m}=\ L_{m}(g_{ij},g_{ij,h},...;y^{\sigma },y_{~j}^{\sigma
},...,y_{~j_{1}...j_{r}}^{\sigma })$ depending on the metric tensor
components and their derivatives up to a certain order $s\in \mathbb{N}$ and
on the $r$-jet of a field $y^{\sigma }$. Typically, in classical general
relativity, $s=0.$

In the case of vacuum Einstein equations%
\begin{equation}
R_{ij}-\dfrac{1}{2}Rg_{ij}=0,  \label{vacuum_efe}
\end{equation}%
the "field components"\ to be varied are the metric tensor components $%
g_{ij} $ (or, more commonly, the inverse metric components $g^{ij}$), hence
the fibered manifold $Y$ is the bundle of metrics $Met(X),$ defined as the
set of symmetric nondegenerate tensors of type (0,2)\ on $X.$ Since both $%
R_{ij}$ and $R$ are of second order in $g_{ij}$, the space we have to work
on is the second order jet bundle $J^{2}Met(X)$.

We denote the local charts on $Met(X)$ by $(V,\psi ),$ with $\psi
=(x^{i},g_{jk})$ and the induced fibered chart on $J^{2}Met(X),$ by $%
(V^{2},\psi ^{2}),$ with $\psi ^{2}=(x^{i},g_{jk};g_{jk,i};g_{jk,il}).$ We
will also use the notations: 
\begin{equation*}
\omega _{jk}=dg_{jk}-g_{jk,i}dx^{i};~\ \omega
_{jk,l}=dg_{jk,l}-g_{jk,li}dx^{i}\ 
\end{equation*}%
for the basic contact forms on $J^{2}Met(X)$. The Riemann tensor, the Ricci
tensor and the Ricci scalar thus become objects on $J^{2}Met(X).$

\subsection{Canonical variational completion of the Ricci tensor}

We will prove in the following that vacuum Einstein equations (\ref%
{vacuum_efe}) can be obtained by means of the canonical variational
completion of the source form with components $R_{ij}.$

Take the following source form on $J^{2}Met(X):$%
\begin{equation}
\varepsilon :=\alpha R^{ij}\sqrt{\left\vert g\right\vert }\omega _{ij}\wedge
\omega _{0},  \label{source_form_Ricci}
\end{equation}%
where $\alpha $ is a (momentarily) arbitrary constant. Its components $%
\varepsilon ^{ij}=\varepsilon ^{ij}(g_{kl};g_{kl,i},g_{kl,ij})$ are given by%
\begin{equation*}
\varepsilon ^{ij}=\alpha R^{ij}\sqrt{\left\vert g\right\vert }.
\end{equation*}

The Vainberg-Tonti Lagrangian $\lambda _{\varepsilon }=\mathcal{L}%
_{\varepsilon }\omega _{0}$ is defined as:%
\begin{equation*}
\mathcal{L}_{\varepsilon }=g_{ij}\underset{0}{\overset{1}{\int }}\varepsilon
^{ij}(ug_{kl};ug_{kl,i};ug_{kl,ij})du.
\end{equation*}

Let us study the behavior of the integrand with respect to homotheties $\chi
_{u}:(g_{kl};g_{kl,i};g_{kl,ij})\mapsto (ug_{kl};ug_{kl,i};ug_{kl,ij}).$
These homotheties induce the transformation $g^{kl}\mapsto u^{-1}g^{kl}$ of
the inverse metric tensor components. The Christoffel symbols%
\begin{equation*}
\Gamma _{~jk}^{i}=\dfrac{1}{2}g^{ih}(g_{hj,k}+g_{hk,j}-g_{jk,h})
\end{equation*}%
are invariant to $\chi _{u}$ and hence the curvature tensor components $%
R_{j~kl}^{~i}=\Gamma _{~jk,l}^{~i}-\Gamma _{~jl,k}^{i}+\Gamma
_{~jk}^{h}\Gamma _{~hl}^{i}-\Gamma _{~jl}^{h}\Gamma _{~hk}^{i}$ are also
invariant. The Ricci tensor $R_{jk}=R_{j~ki}^{~i}$ is obtained just by a
summation process from $R_{j~kl}^{~i}$, which means that it is also
insensitive to $\chi _{u}.$ That is, $R^{ij}=g^{ih}g^{jl}R_{hl}$ will
acquire a $u^{-2}.$

It remains to compute the contribution of $\chi _{u}$ to the factor $\sqrt{%
\left\vert g\right\vert }$. Each line of the matrix $(g_{jk})$ is multiplied
by $u,$ that is, $g=\det (g_{ij})$ will acquire a factor of $u^{4}$ and
finally,%
\begin{equation*}
\sqrt{\left\vert g\circ \chi _{u}\right\vert }=u^{2}\sqrt{\left\vert
g\right\vert }.
\end{equation*}

Substituting into the expression of $\mathcal{L}_{\varepsilon },$ we get
this way,%
\begin{equation*}
\mathcal{L}_{\varepsilon }=g_{ij}\underset{0}{\overset{1}{\int }}u^{0}\alpha
R^{ij}\sqrt{\left\vert g\right\vert }du=\alpha g_{ij}R^{ij}\sqrt{\left\vert
g\right\vert }\underset{0}{\overset{1}{\int }}u^{0}du=\alpha R\sqrt{%
\left\vert g\right\vert }.
\end{equation*}%
Thus, if we choose%
\begin{equation*}
\alpha :=\dfrac{-1}{16\pi \kappa },
\end{equation*}%
the Vainberg-Tonti Lagrangian\textit{\ }$\lambda _{\varepsilon }=\mathcal{L}%
_{\varepsilon }\omega _{0}$ becomes the Hilbert Lagrangian $\lambda _{g}:$%
\begin{equation}
\lambda _{\varepsilon }=\lambda _{g}.  \label{Hilbert_VT_Lagrangian}
\end{equation}

We know, however, that the Euler-Lagrange expressions of $R\sqrt{\left\vert
g\right\vert }$ with respect to $g_{ij}$ are given by (minus) the
contravariant components of the Einstein tensor. In differential form
writing, this is:%
\begin{equation*}
E(\lambda _{\varepsilon })=\dfrac{1}{16\pi \kappa }(R^{ij}-\dfrac{1}{2}%
Rg^{ij})\sqrt{\left\vert g\right\vert }\omega _{ij}\wedge \omega _{0}
\end{equation*}%
hence, we find the variational completion $\tau =E(\lambda _{\varepsilon
})-\varepsilon $ as 
\begin{equation*}
\tau =\dfrac{1}{16\pi \kappa }(2R^{ij}-\dfrac{1}{2}Rg^{ij})\sqrt{\left\vert
g\right\vert }\omega _{ij}\wedge \omega _{0}.
\end{equation*}

\bigskip

\textbf{Remark. }The factor $\alpha $ in (\ref{source_form_Ricci})\ is
actually unessential, the variationally completed equation%
\begin{equation*}
E(\lambda _{\varepsilon })=0
\end{equation*}%
is still the correct vacuum Einstein equation, regardless of its value.

\subsection{Energy-momentum tensors}

Having one term of an energy-momentum tensor, the canonical variational
completion method offers a way of recovering its full expression. We will
apply this method in the case when the known piece is a (non-symmetrized)\
Noether current.

In the case of Einstein equations with matter (\ref{efe}), we will have to
work on a fibered product $Y\times _{X}Met(X)$ over $X$ (where $Y$ is a
fibered manifold with base $X$) with coordinate charts $(V,\psi ),$ $\psi
=(x^{i},y^{\sigma },g_{jk}).$ In this case, one can speak separately about
variations with respect to $y^{\sigma }$ and to $g_{jk}$ and accordingly,
about $Y$-variationality and $Met(X)$-variationality, $Y$- and $Met(X)$%
-variational completions.

Consider a first order Lagrangian $\lambda _{m}$ on $Y\times _{X}Met(X);$ we
suppose in addition that $\lambda _{m}$ does not depend on $x^{j}$ and on
the derivatives $g_{ij,k}.$ Thus, $\lambda _{m}=L_{m}\sqrt{\left\vert
g\right\vert }\omega _{0},$ where%
\begin{equation*}
L_{m}=L_{m}(y^{\sigma },y_{~j}^{\sigma },g_{ij}).
\end{equation*}

In classical relativity theory, there are two major ways of defining
energy-momentum tensors, corresponding to two different contexts:

1) The \textit{canonical energy-momentum tensor}, corresponding to special
relativity\textit{\ }(where $X=\mathbb{R}^{4}$ and the metric tensor is
fixed as $\eta _{ij}=diag(1,-1,-1,-1)$). A Lagrangian $\lambda
_{m}=L_{m}\omega _{0},$ which is invariant to the group of space-time
translations $\tilde{x}^{i}=x^{i}+a^{i},$ $a^{i}=const.,$ gives rise to a
system of conserved Noether currents, called the canonical energy-momentum
tensor (invariance to space-time translations amounts to the above
assumption that $L_{m}$ does not explicitly depend on $x^{i}$). These
Noether currents are given by, \cite{Landau}:%
\begin{equation}
\tilde{T}^{ij}=\eta ^{ik}(y_{~,k}^{\sigma }\dfrac{\partial L_{m}}{\partial
y_{~,j}^{\sigma }}-\delta _{k}^{j}L_{m}).  \label{Noether_current_general}
\end{equation}

The canonical energy-momentum tensor $\tilde{T}^{ij}$ is, generally, not
symmetric - which is inconvenient, since symmetry is required on physical
grounds (angular momentum conservation). This is usually solved by adding a
divergence-free term, thus obtaining a tensor $\overset{can}{T}\overset{}{%
^{ij}}$ which is symmetric and still conserved, i.e., $\overset{can}{T}%
\overset{}{_{~~,j}^{ij}}=0.$ There are multiple possibilities of choosing
the symmetrization term, \cite{Forger}.

2) In general relativity\textit{\ }(where $(X,g_{ij})$ is an arbitrary
Lorentzian manifold), energy-momentum tensors (\textit{Hilbert, or metric
energy-momentum tensors}) $\overset{met}{T}\overset{}{^{ij}}$ are defined by
means of functional derivatives of the matter Lagrangian $\lambda _{m}=%
\mathcal{L}_{m}\omega _{0},$ $\mathcal{L}_{m}=L_{m}\sqrt{\left\vert
g\right\vert },$ with respect to $g_{ij}$:%
\begin{equation}
-\dfrac{1}{2}\mathcal{T}^{ij}:=\dfrac{\delta \mathcal{L}_{m}}{\delta g_{ij}}%
;~\ \ \overset{met}{T}\overset{}{^{ij}}:=\dfrac{1}{\sqrt{\left\vert
g\right\vert }}\mathcal{T}^{ij}.  \label{def_sem_tensor_gr}
\end{equation}%
Here, $L_{m}=L_{m}(y^{\sigma },y_{~j}^{\sigma },y_{~j}^{\sigma };g_{ij})$ is
a differential invariant (a "\textit{scalar}"), hence the Lagrangian $%
\lambda _{m}$ is invariant to (transformations on $J^{r}Y$ induced by)
arbitrary diffeomorphisms on $X$. As a result, $\overset{met}{T}\overset{}{%
^{ij}}$ obeys on-shell the \textit{covariant conservation law }$\overset{met}%
{T}\overset{}{_{~~;j}^{ij}}=0$ and also, has gauge invariance properties, 
\cite{Forger}. Moreover, $\overset{met}{T}\overset{}{^{ij}}$ is, by
construction, symmetric.

The two procedures of defining the energy-momentum tensor are fundamentally
different and obviously require a thorough geometric analysis. Just as a
first remark, they generally do not even make sense at the same time:\ in
special relativity, where the metric is fixed, it makes no sense to speak
about variations of a Lagrangian with respect to the metric. On the other
hand, in general relativity, where $X$ is an arbitrary manifold, space-time
translations $\tilde{x}^{i}=x^{i}+a^{i},$ $a^{i}=const.,$ cannot be defined
geometrically. However, there is a realm (see, e.g., \cite{Leclerc}) where
both procedures can be applied, namely, when:%
\begin{equation}
X=\mathbb{R}^{4},~\ g_{ij}-arbitrary  \label{border_zone}
\end{equation}%
(actually, in \cite{Leclerc}, it is pointed out the particular case of 
\textit{weak metrics --} in which the author studies the equivalence between
the two definitions. Still, for our purposes, we do not need the assumption
that the metric is weak).

\bigskip

For a special-relativistic Lagrangian%
\begin{equation}
\lambda _{m}=L_{m}\omega _{0},~~L_{m}=L_{m}(y^{\sigma },y_{~i}^{\sigma
},g_{ij}=\eta _{ij}),  \label{special_rel_L}
\end{equation}%
the canonical variational completion offers a recipe of symmetrization of
the Noether current $\tilde{T}^{ij}$. We will do this in three steps:

\textbf{Step 1. }We leave for the moment the special relativistic context
and formally allow $g^{ij}$ to vary. Abiding by the principle of general
covariance, \cite{Landau}, a straightforward generalization of (\ref%
{Noether_current_general})\ to the new context is given by the tensor
density:%
\begin{equation}
\mathcal{\tilde{T}}^{ij}=g^{ik}(y_{~;k}^{\sigma }\dfrac{\partial L_{m}}{%
\partial y_{~;j}^{\sigma }}-\delta _{k}^{j}L_{m})\sqrt{\left\vert
g\right\vert },  \label{curved_space_Noether_current}
\end{equation}%
where the semicolon $_{;k}$ denotes (formal)\ covariant differentiation with
respect to $\partial /\partial x^{k}$.

\textit{Note: }In the above, $y^{\sigma }$ are tensors of some unspecified
rank (the upper position of the index is chosen just for convenience; $%
y^{\sigma }$ can very well be components of, e.g., a scalar, a covector
field or of a tensor of type (0,2)).

\bigskip

\textbf{Step 2. }Taking into account (\ref{def_sem_tensor_gr}), we consider
the source form $\varepsilon =\alpha \mathcal{\tilde{T}}^{ij}\omega
_{ij}\wedge \omega _{0}$ on $J^{1}(Y\times _{X}Met(X)),$ with components$\
\varepsilon ^{ij}=\alpha \mathcal{\tilde{T}}^{ij}(y^{\sigma },y_{~j}^{\sigma
},g_{kh})$, where $\alpha \in \mathbb{R}$ is a constant. Its $Met(X)$%
-Vainberg-Tonti Lagrangian $\lambda _{\varepsilon }:=\mathcal{L}%
_{\varepsilon }\omega _{0}$ is:%
\begin{equation*}
\mathcal{L}_{\varepsilon }=\alpha g_{ij}\underset{0}{\overset{1}{\int }}(%
\mathcal{\tilde{T}}^{ij}\circ \chi _{u})du,
\end{equation*}%
where $\chi _{u}(y^{\sigma },y_{~i}^{\sigma },g_{ij}):=(y^{\sigma
},y_{~i}^{\sigma },ug_{ij})$ only affects the metric components.
Substituting $\mathcal{\tilde{T}}^{ij}$ from (\ref%
{curved_space_Noether_current}) and\ taking into account that $\chi _{u}$
leaves Christoffel symbols invariant and that $\delta _{i}^{i}=\dim (X)=4,$
we have: 
\begin{equation}
\mathcal{L}_{\varepsilon }=\alpha \underset{0}{\overset{1}{\int }}%
u(y_{~;i}^{\sigma }\dfrac{\partial (L_{m}\circ \chi _{u})}{\partial
y_{~;i}^{\sigma }}-4L_{m}\circ \chi _{u})\sqrt{|g|}du.
\label{VT_Lagrangian_canonical_em_tensor}
\end{equation}%
Further, we calculate the Hilbert energy-momentum tensor of $\lambda
_{\varepsilon }$ as:%
\begin{equation}
-\dfrac{1}{2}\overset{met}{T}\overset{}{^{ij}}:=\dfrac{1}{\sqrt{\left\vert
g\right\vert }}\dfrac{\delta \mathcal{L}_{\varepsilon }}{\delta g_{ij}}.
\label{symmetrized_Noether_current}
\end{equation}%
\bigskip

\textbf{Step 3. }Finally, particularize in (\ref{symmetrized_Noether_current}%
) $g_{ij}$ as $\eta _{ij}\ $and define%
\begin{equation*}
\overset{met}{T}\overset{}{^{ij}}:|_{g_{ij}=\eta _{ij}}=:T^{ij}.
\end{equation*}%
This way, $\overset{met}{T}\overset{}{^{ij}}$ is defined up to
multiplication by the constant $\alpha $. This constant can then be
adjusted, for instance, in such a way that the obtained symmetrization term%
\begin{equation}
\tau ^{ij}:=T^{ij}-\tilde{T}^{ij}  \label{general_correction_em_tensor}
\end{equation}%
is independent from $\tilde{T}^{ij}$ (it does not contain any multiple of $%
\tilde{T}^{ij}$).

The covariant conservation law of $\overset{met}{T}\overset{}{^{ij}}$
(obtained as a consequence of the fact that $\overset{met}{T}\overset{}{^{ij}%
}$ is a Hilbert energy-momentum tensor)\ now transforms into the usual
conservation law: $T_{~~,j}^{ij}=0.$ Thus, the obtained energy-momentum
tensor $T_{ij}$ is, as required, both symmetric and conserved. Moreover, the
symmetrzation term $\tau ^{ij}$ offers a measure of the non-$Met(X)$%
-variationality of $\tilde{T}^{ij}.$

\bigskip

\textbf{Example:\ }\textit{energy-momentum tensor of the electromagnetic
field.}\newline
The electromagnetic field is described by the potential 1-form $%
A=A_{i}dx^{i} $ on $X$ and by the 2-form\textit{\ }$F:=dA=\dfrac{1}{2}%
F_{ij}dx^{i}\wedge dx^{j}.$

In the special relativistic case $g_{ij}=\eta _{ij},$ we have $%
F_{ij}=A_{j,i}-A_{i,j},$ or, in terms of the contravariant components $%
A^{i}: $ $F_{ij}=\eta _{jk}A_{~,i}^{k}-\eta _{ik}A_{~,j}^{k}.$ The
Lagrangian of the electromagnetic field is $\lambda _{f}=L_{f}\omega _{0}$
with%
\begin{equation}
L_{f}=-\dfrac{1}{16\pi }F_{ij}F^{ij};  \label{em_field_Lagrangian}
\end{equation}%
Translational invariance of $\lambda _{f}$ leads to the Noether current, 
\cite{Landau}:%
\begin{equation}
\tilde{T}^{ij}=-\dfrac{1}{4\pi }\eta ^{ih}\dfrac{\partial A^{l}}{\partial
x^{h}}F_{~l}^{j}+\dfrac{1}{16\pi }\eta ^{ij}F_{kl}F^{kl}.
\label{Noether_current_em_field}
\end{equation}

The curved space generalization of $\tilde{T}^{ij}$ in (\ref%
{Noether_current_em_field}) is the tensor density:

\begin{equation}
\mathcal{\tilde{T}}^{ij}=(-\dfrac{1}{4\pi }g^{ih}A_{~;h}^{l}F_{~l}^{j}+%
\dfrac{1}{16\pi }g^{ij}F_{kl}F^{kl})\sqrt{\left\vert g\right\vert }
\label{Noether_current_contravariant}
\end{equation}%
where, this time: 
\begin{equation}
F_{ij}=g_{jk}A_{~;i}^{k}-g_{ik}A_{~;j}^{k}.  \label{Faraday_form}
\end{equation}%
Further, we calculate the Vainberg-Tonti Lagrangian of the source form%
\begin{equation*}
\varepsilon =\alpha \mathcal{\tilde{T}}^{ij}\omega _{ij}\wedge \omega _{0},
\end{equation*}%
where $\mathcal{\tilde{T}}^{ij}=\mathcal{\tilde{T}}%
^{ij}(A^{k};A_{,l}^{k};g_{kl};g_{kl,h})$. We prefer to use $A^{k}$ rather
than $A_{k}:=g_{kl}A^{l}$ as the field variables for a reason which will
become transparent below. This way, $\chi _{u}$ acts as follows:%
\begin{equation*}
g_{ij}\circ \chi _{u}=ug_{ij},~\ g^{ij}\circ \chi _{u}=u^{-1}g^{ij},
\end{equation*}%
while $\chi _{u}$ does not affect the field variables $y^{\sigma }=A^{k}$.
Again, the Christoffel symbols $\Gamma _{~jk}^{i}$ are invariant to $\chi
_{u}.$ Expressing $F_{ij}$ as in (\ref{Faraday_form}), we can now determine
the effect of $\chi _{u}$ on each term of $\mathcal{\tilde{T}}^{ij}:$%
\begin{equation*}
A_{~;i}^{l}\circ \chi _{u}=A_{~;i}^{l};~\ F_{jl}\circ \chi _{u}=uF_{jl};~\ \
F^{kl}\circ \chi _{u}=u^{-1}F^{kl},~\ \ \sqrt{\left\vert g\right\vert }\circ
\chi _{u}=u^{2}\sqrt{\left\vert g\right\vert }.
\end{equation*}%
All in all, we have:%
\begin{equation*}
\mathcal{\tilde{T}}^{ij}\circ \chi _{u}=u\mathcal{\tilde{T}}^{ij}
\end{equation*}%
and hence, the Vainberg-Tonti Lagrangian $\lambda _{\varepsilon }=\mathcal{L}%
_{\varepsilon }\omega _{0}$ is given by:%
\begin{equation*}
\mathcal{L}_{\varepsilon }=g_{ij}\underset{0}{\overset{1}{\int }}u\alpha 
\mathcal{\tilde{T}}^{ij}du=\dfrac{\alpha }{2}g_{ij}\mathcal{\tilde{T}}^{ij},
\end{equation*}%
that is, 
\begin{equation}
\mathcal{L}_{\varepsilon }=\alpha (-\dfrac{1}{8\pi }A^{l;k}F_{kl}+\dfrac{1}{%
8\pi }F_{kl}F^{kl})\sqrt{\left\vert g\right\vert }.
\label{em_Lagrangian_rough}
\end{equation}

Taking into account that $F_{kl}=-F_{lk},$ the term $A^{l;k}F_{kl}$ in the
above can be re-expressed as:\ $A^{l;k}F_{kl}=\dfrac{1}{2}%
(A^{l;k}-A^{k;l})F_{kl}=\dfrac{1}{2}F^{kl}F_{kl};$ substituting into (\ref%
{em_Lagrangian_rough}), we finally obtain the $Met(X)$-Vainberg-Tonti
Lagrangian of (\ref{Noether_current_contravariant}) as:%
\begin{equation}
\lambda _{\varepsilon }=\dfrac{\alpha }{16\pi }F^{kl}F_{kl}\sqrt{\left\vert
g\right\vert }\omega _{0}=-\alpha \lambda _{f}.
\label{VT_Lagrangian_em_field}
\end{equation}

But, variation of $\lambda _{f}$ with respect to $g^{ij}$ is well-known, 
\cite{Landau}; namely, we will get for $\lambda _{\varepsilon }=-\alpha
\lambda _{f}$ the Hilbert energy-momentum tensor%
\begin{equation*}
\overset{met}{T}\overset{}{^{ij}}(\alpha )=-\alpha (-\dfrac{1}{4\pi }%
F^{il}F_{~l}^{j}+\dfrac{1}{16\pi }g^{ij}F_{kl}F^{kl}).
\end{equation*}

Particularizing now $g^{ij}=\eta ^{ij},$ we get the symmetrized
energy-momentum tensor:\ 
\begin{equation*}
T^{ij}:=\overset{met}{T}\overset{}{^{ij}}=-\alpha (\tilde{T}^{ij}+\dfrac{1}{%
4\pi }A_{~,l}^{i}F^{jl}).
\end{equation*}
Taking $\alpha :=-1$ (which provides $\overset{met}{T}\overset{}{^{ij}}=%
\tilde{T}^{ij}+independent\_term$), we obtain $\lambda _{\varepsilon
}=\lambda _{f}$ and the symmetrization term:%
\begin{equation*}
\tau ^{ij}=\dfrac{1}{4\pi }A_{~,l}^{i}F^{jl},
\end{equation*}%
or, in covariant writing, 
\begin{equation*}
\tau _{ij}=\dfrac{1}{4\pi }A_{i,l}F_{j}^{~l}
\end{equation*}%
as the correction term. This is the classical symmetrization term, \cite%
{Landau}, yet, obtained here by a completely different reasoning.

\bigskip

\textbf{Remarks.}

1)\ If, for a given (symmetrized or not)\ energy-momentum tensor $\tilde{T}%
^{ij}$, the Lagrangian $\lambda _{m}$ is not known, a Lagrangian can be
constructed as the $Met(X)$-Vainberg-Tonti Lagrangian (\ref%
{VT_Lagrangian_canonical_em_tensor}); if a Lagrangian $\lambda _{m}$ is
already known, the above gives an alternative construction.

1)\ If the given matter Lagrangian density $\mathcal{L}_{m}$ is \textit{%
homogeneous} both in the metric components and in the derivatives $%
y_{~;i}^{\sigma }$ (the homogeneity degrees need not coincide) then,
applying Euler's theorem in (\ref{VT_Lagrangian_canonical_em_tensor}), we
see that the Vainberg-Tonti Lagrangian density $\mathcal{L}_{\varepsilon }$
in (\ref{VT_Lagrangian_canonical_em_tensor}) actually coincides, up to
multiplication by some constant, with the matter Lagrangian density $%
\mathcal{L}_{m}.$ In this case, we can always choose $\alpha $ such that $%
\mathcal{L}_{\varepsilon }=\mathcal{L}_{m}.$ In this case, the
symmetrization term coincides with the one in \cite{Gotay}, yet, it is found
just by considerations of variationality.

2) If we had worked with the potential 1-form components $A_{i}$ (instead of
the vector field components $A^{i}$) as our field variables, we would have
had $F_{ij}=A_{j;i}-A_{i;j}$ - invariant to $\chi _{u}$ and by a similar
reasoning to the above, we would have got $\mathcal{\tilde{T}}%
^{ij}(ug_{kl})=u^{-1}\mathcal{\tilde{T}}^{ij}(g_{kl})$ and, consequently, to 
$\mathcal{L}_{\varepsilon }=(g^{ij}\mathcal{\tilde{T}}_{ij})\underset{0}{%
\overset{1}{\int }}u^{-1}du.$ But, since the latter integral does not have a
finite value, we could not have calculated $\mathcal{L}_{\varepsilon }$ this
way. Hence, it appears that, at least in this case, the 4-potential \textit{%
vector field} components $A^{i}$ are a more advantageous choice for our
dynamical variables.

\section{An example in first order mechanics}

Take $Y=\mathbb{R}\times \mathbb{R}^{n}$, with local coordinates $%
(t,q^{\sigma });$ on the second jet prolongation $J^{2}Y,$ we denote the
induced local coordinates by $(t,q^{\sigma },\dot{q}^{\sigma },\ddot{q}%
^{\sigma }).$

Consider the second order source form%
\begin{equation*}
\varepsilon =\varepsilon _{\sigma }\omega ^{\sigma }\wedge dt,
\end{equation*}%
\begin{equation}
\varepsilon _{\sigma }=m_{\sigma \nu }\ddot{q}^{\nu }+k_{\sigma \nu }q^{\nu
}+\dfrac{\partial F}{\partial \dot{q}^{\sigma }},
\label{damped_oscill_epsilon}
\end{equation}%
where:

- $m_{\sigma \nu },$ $k_{\sigma \nu }$ are constant and symmetric;

- $F=F(\dot{q}^{\sigma })$ is homogeneous of some degree $p\geq 1$ in $\dot{q%
}^{\sigma }.$

\bigskip

The ODE system $\varepsilon _{\sigma }=0$ is generally non-variational. Let
us determine its canonical variational completion. The Vainberg-Tonti
Lagrangian attached to $\varepsilon $ is $\lambda _{\varepsilon }=\mathcal{L}%
_{\varepsilon }dt,$ with%
\begin{equation*}
\mathcal{L}_{\varepsilon }=q^{\sigma }\overset{1}{\underset{0}{\int }}%
\varepsilon _{\sigma }(t,uq^{\nu },u\dot{q}^{\nu },u\ddot{q}^{\nu
})du=q^{\sigma }\overset{1}{\underset{0}{\int }}(m_{\sigma \nu }u\ddot{q}%
^{\nu }+k_{\sigma \nu }uq^{\nu }+\dfrac{\partial F}{\partial \dot{q}^{\sigma
}}(u\dot{q}^{\nu }))du.
\end{equation*}%
Taking into account the homogeneity degree of $F,$ this is:%
\begin{eqnarray*}
\mathcal{L}_{\varepsilon } &=&q^{\sigma }\overset{1}{\underset{0}{\int }}%
[u(m_{\sigma \nu }\ddot{q}^{\nu }+k_{\sigma \nu }q^{\nu })+u^{p-1}\dfrac{%
\partial F}{\partial \dot{q}^{\sigma }}]du= \\
&=&\dfrac{1}{2}(m_{\sigma \nu }\ddot{q}^{\nu }q^{\sigma }+k_{\sigma \nu
}q^{\sigma }q^{\nu })+\dfrac{1}{p}q^{\sigma }\dfrac{\partial F}{\partial 
\dot{q}^{\sigma }}.
\end{eqnarray*}%
The term $\dfrac{1}{2}m_{\sigma \nu }\ddot{q}^{\nu }q^{\sigma }$ differs by
a total derivative $d_{t}(\dfrac{1}{2}m_{\sigma \nu }\dot{q}^{\nu }q^{\sigma
})$ from $-\dfrac{1}{2}m_{\sigma \nu }\dot{q}^{\nu }\dot{q}^{\sigma },$
hence the two expressions are dynamically equivalent. We will thus prefer to
take the latter, which is of lower order and thus, we obtain the following
Lagrangian function, which is equivalent to the Vainberg-Tonti one:%
\begin{equation}
\mathcal{L}=\dfrac{1}{2}(-m_{\sigma \nu }\dot{q}^{\nu }\dot{q}^{\sigma
}+k_{\sigma \nu }q^{\sigma }q^{\nu })+\dfrac{1}{p}q^{\sigma }\dfrac{\partial
F}{\partial \dot{q}^{\sigma }}.  \label{VT_Lagrangian_oscillations}
\end{equation}

Let us determine the Euler-Lagrange form of $\mathcal{L}.\ $We have, on one
hand:%
\begin{equation*}
\dfrac{\partial \mathcal{L}}{\partial q^{\rho }}=k_{\sigma \rho }q^{\sigma }+%
\dfrac{1}{p}\dfrac{\partial F}{\partial \dot{q}^{\rho }}
\end{equation*}%
and, on the other hand,%
\begin{eqnarray*}
\dfrac{\partial \mathcal{L}}{\partial \dot{q}^{\rho }} &=&-m_{\sigma \rho }%
\dot{q}^{\rho }+\dfrac{1}{p}\dfrac{\partial ^{2}F}{\partial \dot{q}^{\sigma
}\partial \dot{q}^{\rho }}q^{\sigma },~\ \ \ \  \\
d_{t}(\dfrac{\partial \mathcal{L}}{\partial \dot{q}^{\rho }}) &=&-m_{\sigma
\rho }\ddot{q}^{\rho }+\dfrac{1}{p}\dfrac{\partial ^{3}F}{\partial \dot{q}%
^{\sigma }\partial \dot{q}^{\rho }\partial \dot{q}^{\nu }}\ddot{q}^{\nu
}q^{\sigma }+\dfrac{1}{p}\dfrac{\partial ^{2}F}{\partial \dot{q}^{\sigma
}\partial \dot{q}^{\rho }}\dot{q}^{\sigma };
\end{eqnarray*}%
taking again into account that $\dfrac{\partial F}{\partial \dot{q}^{\rho }}$
is homogeneous of degree $p-1,$ the latter term is: $\dfrac{1}{p}\dfrac{%
\partial ^{2}F}{\partial \dot{q}^{\sigma }\partial \dot{q}^{\rho }}\dot{q}%
^{\sigma }=\dfrac{p-1}{p}\dfrac{\partial F}{\partial \dot{q}^{\rho }}$ and,
finally, 
\begin{equation*}
E_{\rho }(\mathcal{L})=(m_{\sigma \rho }\ddot{q}^{\sigma }+k_{\sigma \rho
}q^{\sigma })+\dfrac{2-p}{p}\dfrac{\partial F}{\partial \dot{q}^{\rho }}-%
\dfrac{1}{p}\dfrac{\partial ^{3}F}{\partial \dot{q}^{\sigma }\partial \dot{q}%
^{\rho }\partial \dot{q}^{\nu }}\ddot{q}^{\nu }q^{\sigma }.
\end{equation*}%
We find the variational completion $\tau =\tau _{\rho }(t,q,\dot{q})\omega
^{\rho }\wedge dt$ as:%
\begin{equation}
\tau _{\rho }=2(\dfrac{1}{p}-1)\dfrac{\partial F}{\partial \dot{q}^{\rho }}-%
\dfrac{1}{p}\dfrac{\partial ^{3}F}{\partial \dot{q}^{\sigma }\partial \dot{q}%
^{\rho }\partial \dot{q}^{\nu }}\ddot{q}^{\nu }q^{\sigma }.
\label{dissipative_force_correction}
\end{equation}

\bigskip

\textbf{Particular cases:}\textit{\ }

1)\ If $F=0,$ the system $\varepsilon _{\sigma }=0$ is equivalent to%
\begin{equation*}
m_{\sigma \nu }\ddot{q}^{\nu }+k_{\sigma \nu }q^{\nu }=0.
\end{equation*}%
These equations characterize free small oscillations with multiple degrees
of freedom, \cite{Landau}. They are known to be variational; their
Lagrangian function $\mathcal{L}=\dfrac{1}{2}(-m_{\sigma \nu }\dot{q}^{\nu }%
\dot{q}^{\sigma }+k_{\sigma \nu }q^{\sigma }q^{\nu })$ coincides (as
expected), with (\ref{VT_Lagrangian_oscillations}).

2)\ If $p=2$ and $F$ is quadratic in $\dot{q}$: 
\begin{equation*}
F=\dfrac{1}{2}\alpha _{\sigma \nu }\dot{q}^{\sigma }\dot{q}^{\nu },
\end{equation*}%
(where $\alpha _{\sigma \nu }=\alpha _{\nu \sigma }\in \mathbb{R}$) the ODE
system $\varepsilon _{\sigma }=0$ characterizes, \cite{Landau-mech}, Section
25, linearly damped oscillations. In this case, the function $F$ is called
the \textit{Rayleigh dissipation function} and is interpreted as the rate of
energy dissipation in the system. In (\ref{damped_oscill_epsilon}), the last
term (with a minus in front) $-\dfrac{\partial F}{\partial \dot{q}^{\sigma }}%
=-a_{\sigma \nu }\dot{q}^{\nu }$ is interpreted as a friction force. In this
case, the canonical variational completion (\ref%
{dissipative_force_correction}) is given by%
\begin{equation*}
\tau _{\rho }=-\dfrac{\partial F}{\partial \dot{q}^{\rho }}
\end{equation*}%
and the variationally completed\ equations are:%
\begin{equation}
m_{\rho \nu }\ddot{q}^{\nu }+k_{\rho \nu }q^{\nu }=0,
\label{free_oscillations}
\end{equation}%
which are precisely the equations of "undamped"\ oscillations. That is, the
friction force $\dfrac{\partial F}{\partial \dot{q}^{\rho }}$ has, in this
case, the meaning of obstruction from variationality\ of the equations.

\textbf{Remark. }In other cases (e.g., when $-\dfrac{\partial F}{\partial 
\dot{q}^{\rho }}$ is quadratic or cubic in $\dot{q}^{\sigma }$), the
variationally completed equations will not coincide anymore with the
equations (\ref{free_oscillations}) of undamped oscillations.

\textbf{Acknowledgment. }The work was partially supported by the Grant of
Transilvania University 2013 (\textit{Bursa Universitatii "Transilvania"\
2013}).


\begin{thebibliography}{99}
\bibitem{Anderson} I.M. Anderson, \textit{Aspects of the inverse problem to
the calculus of variations}, Archivum Mathematicum, 24 (4) (1988), 181--202.

\bibitem{Forger} M. Forger, H. Romer, \textit{Currents and the energy
momentum tensor in classical field theory: A fresh look at an old problem} -
Annals Phys. 309 (2004).

\bibitem{Gotay} M.J. Gotay, J. E. Marsden, Stress-Energy-Momentum tensors
and the Belinfante-Rosenfeld formula, Contemp. Math. 132 (1992), 367-392.

\bibitem{Grigore} D.R. Grigore, \textit{On an order reduction theorem in the
Lagrangian formalism, }Il Nuovo Cimento B, 111(12) 1996, 1439-1447.

\bibitem{Krupka-natural-L} D. Krupka, \textit{Natural Lagrangian structures}%
, Differential Geometry, Banach Center Publ. 12, Diff. Geom. Semester,
Warsaw, Sept.-Dec. 1979; Polish Scientific Publ., Warsaw, 1984, 185-210.

\bibitem{Krupka-local-structure-L} D. Krupka, \textit{On the local structure
of the Euler-Lagrange mapping of the calculus of variations}, Proc. Conf. on
Diff. Geom. Appl., Univerzita Karlova, Czech Republic, 1981, 181-188, arXiv:
math-ph/0203034.

\bibitem{Krupka-VT-Lagrangian} D. Krupka, \textit{The Vainberg-Tonti
Lagrangian and the Euler-Lagrange mapping}, in: F. Cantrijn, B. Langerock,
Eds., Differential Geometric Methods in Mechanics and Field Theory, Volume
in Honour of W. Sarlet, Gent, Academia Press, 2007, 81-90.

\bibitem{Landau} L.D. Landau, E. M. Lifschiz, \textit{The Classical Theory
of Fields}, 4th edn., Elsevier, 1975.

\bibitem{Landau-mech} L.D. Landau, E. M. Lifschiz, \textit{Mechanics}, 3rd
edn., Elsevier, 1976.

\bibitem{Leclerc} M. Leclerc, \textit{Canonical and gravitational
stress-energy tensors, }Int. J. Mod. Phys. D15 (2006) 959-990.
\end{thebibliography}
\end{document}